\RequirePackage{fix-cm}
\documentclass[smallextended]{svjour3}       
\smartqed  
\usepackage{graphicx}
\usepackage[utf8]{inputenc}
\usepackage{todonotes}
\usepackage{tikz}
\usepackage[font=small,labelfont=bf,tableposition=top]{caption}
\usepackage[T1]{fontenc}
\usepackage{subcaption}
\usepackage{amsmath,amssymb,amsfonts}
\usepackage{graphicx}
\usepackage{color}
\usepackage{textcomp}
\usepackage{listings}
\usepackage{etoolbox}

\usepackage{graphicx}
\usepackage{multirow}
\usepackage{colortbl}
\usepackage{framed}
\usepackage{tabularx}
\usepackage[ruled,noend,linesnumbered]{algorithm2e} 
\usepackage{multirow,bigdelim}
\usepackage{enumitem}
\renewcommand{\descriptionlabel}[1]{\hspace{\labelsep}\textbf{#1}}
\usepackage{balance}
\usepackage{url}
\usepackage{booktabs}
\usepackage{threeparttable}
\usepackage{natbib}
\usepackage{authblk}
\usepackage[breaklinks,colorlinks,citecolor=blue]{hyperref}

\makeatletter
\let\old@lstKV@SwitchCases\lstKV@SwitchCases
\def\lstKV@SwitchCases#1#2#3{}
\makeatother
\usepackage{lstlinebgrd}
\makeatletter
\let\lstKV@SwitchCases\old@lstKV@SwitchCases

\lst@Key{numbers}{none}{%
    \def\lst@PlaceNumber{\lst@linebgrd}%
    \lstKV@SwitchCases{#1}%
    {none:\\%
     left:\def\lst@PlaceNumber{\llap{\normalfont
                \lst@numberstyle{\thelstnumber}\kern\lst@numbersep}\lst@linebgrd}\\%
     right:\def\lst@PlaceNumber{\rlap{\normalfont
                \kern\linewidth \kern\lst@numbersep
                \lst@numberstyle{\thelstnumber}}\lst@linebgrd}%
    }{\PackageError{Listings}{Numbers #1 unknown}\@ehc}}
\makeatother
\newcommand{\faultloc}{query-based fault localization}
\newcommand{\faultloccap}{Query-based fault localization}
\newcommand{\cofind}{collaborative bug finding}
\newcommand{\rqcat}{RQ2}
\newcommand{\studyrepair}{30\%}
\newcommand{\studycontext}{37\%}
\newcommand{\cofix}{collaborative bug fixing}

\newcommand{\precfix}{\textsc{Precfix}} 
\newcommand{\nav}{navigator}
\newcommand{\numstudy}{150} 
\newcommand{\numstudyproj}{150} 
\newcommand{\numevaljava}{152}
\newcommand{\numevaljavaproj}{25}

\newcommand{\numjavarel}{75}
\newcommand{\numandrel}{25}

\newcommand{\numevalandproj}{11}
\newcommand{\numevaland}{97}
\newcommand{\numiss}{10} 
\newcommand{\numeval}{249} 
\newcommand{\numreplypos}{six} 
\newcommand{\numevalreply}{seven} 
\newcommand{\numevalconfirm}{three} 
\newcommand{\numevalclose}{three}
\newcommand{\preckoneavgjava}{0.3399}
\newcommand{\preckoneavgand}{0.4433}
\newcommand{\preckoneavgjavagit}{0.3333}

\newcommand{\numcomment}{36}
\newcommand{\nummytoolrep}{9}
\newcommand{\nummytoolcon}{20}

\newcommand{\numop}{five}

\newcommand{\mytool}{\textsc{CrossFix}} 
\newcommand{\Plus}{\mathord{\begin{tikzpicture}[baseline=0ex, line width=0.5, scale=0.06]\draw (1,0) -- (1,2);
\draw (0,1) -- (2,1);
\end{tikzpicture}}}
\newcommand{\Minus}{\mathord{\begin{tikzpicture}[baseline=0ex, line width=0.5, scale=0.06]
\draw (0,1) -- (2,1);
\end{tikzpicture}}}

\usepackage{etoolbox}
\newcommand*{\affaddr}[1]{#1} 
\newcommand*{\affmark}[1][*]{\textsuperscript{#1}}

\journalname{Empirical Software Engineering}

\begin{document}

\title{\mytool: Collaborative bug fixing by recommending similar bugs
}


\author{Shin Hwei Tan        \and
        Ziqiang Li \affmark[*] \and
        Lu Yan
}


\institute{
 \affaddr{\affmark[*] Shin Hwei Tan (Corresponding author)}  \at
              Southern University of Science and Technology,\\
              Department of Computer Science and Engineering, Shenzhen, China\\
              \email{tansh3@sustech.edu.cn}  \\
              ORCID: 0000-0001-8633-3372
        \and
            \affaddr{\affmark[*] Ziqiang Li (Joint first authors)}\at
            Southern University of Science and Technology, \\
            Department of Computer Science and Engineering, Shenzhen, China\\
            \email{lizq2019@mail.sustech.edu.cn}
        \and
            Lu Yan \at
            Southern University of Science and Technology, \\
            Department of Computer Science and Engineering, Shenzhen, China\\
            \email{lunaryan1998@gmail.com}
}

\date{Received: date / Accepted: date}

\maketitle

\begin{abstract}
Many automated program repair techniques have been proposed for fixing bugs. Some of these techniques use the information beyond the given buggy program and test suite to improve the quality of generated patches. However, there are several limitations that hinder the wide adoption of these techniques, including (1) they rely on a fixed set of repair templates for patch generation or reference implementation, (2) searching for the suitable reference implementation is challenging, (3) generated patches are not explainable. Meanwhile, a recent approach shows that similar bugs exist across different projects and one could use the GitHub issue from a different project for finding new bugs for a related project. We propose \cofix{}, a novel approach that suggests bug reports that describe a similar bug. Our study redefines similar bugs as bugs that share the (1) same libraries, (2) same functionalities, (3) same reproduction steps, (4) same configurations, (5) same outcomes, or (6) same errors. Moreover, our study revealed the usefulness of similar bugs in helping developers in finding more context about the bug and fixing. Based on our study, we design \mytool{}, a tool that automatically suggests relevant GitHub issues based on an open GitHub issue. Our evaluation on \numeval{} open issues from Java and Android projects shows that \mytool{} could suggest similar bugs to help developers in debugging and fixing. 
\keywords{Debugging and program repair \and Collective Knowledge \and Mining Software Repository}
\end{abstract}

\section{Introduction}
\label{sec:intro}

In recent years, many automated program repair techniques have been proposed to assist developers in reducing the manual efforts in localizing and fixing bugs~\citep{xiong2017precise,genprogae,rsrepair,nguyen2013semfix,kim2013automatic,angelix,prophet}. Given a buggy program $P$ and a test-suite $T$ with at least one failing test, these approaches automatically generate a modified program $P'$ which passes all tests in $T$. Despite recent advancement in automated program repair techniques, these techniques have not been widely adopted due to several limitations. First, the quality of the automatically generated patches program repair techniques relies heavily on the quality of the provided test suite and the generated patches may overfit the test suite~\citep{cure,yi2018correlation}. Second, as existing repair techniques usually rely on a set of predefined patterns~\citep{kim2013automatic,tan2016anti} or modifications of the buggy program~\citep{ghanbari2019practical,DBLP:conf/icst/DebroyW10,genprog}, they are limited to fixing defects that requires modifies one or two lines of code, and are more likely to fix easy defects rather than hard ones~\citep{motwani2018automated}. Third, automated repair techniques mostly treat the given test suite as blackbox and do not consider the target defect classes that these techniques aim to address~\citep{critical-review}. Hence, most automatically generated patches may be incorrect~\citep{patchplausibility}. Fourth, the intention of the automatically generated patches may be difficult to understand. In fact, a recent article suggested presenting hints to developers in natural language to explain the generated patches~\citep{goues2019automated}.

With the increasing popularity of open-source software repositories like GitHub, bug reports~\footnote{We use the terms ``bug reports'' and ``GitHub issues'' interchangeably} for many open source projects are made publicly available. Bug reports may contain a wealth of information that could assist developers in localizing and fixing bugs more efficiently~\citep{yang2016combining,wang2008approach,rocha2016empirical}. To exploit the redundancies of bug reports in open source repositories, recent research on \emph{collaborative bug finding} uses the similarities between Android apps for recommending relevant bug reports for a given app under test~\citep{collabbug}. This study revealed several interesting observations that may help in solving the aforementioned problems in automated program repair: (1) similar bugs could exist even across different applications, and (2) one could treat another similar application as a competent pair programmer who can help in discovering new bugs.  

\begin{table*}[t]
\centering
 \caption{Maui's developer refers to a similar bug in deeplearning4j}
\label{tab:similar-bug}

 \resizebox{\linewidth}{!} {

\begin{tabular}{l|l}
\multicolumn{1}{c}{} &\multicolumn{1}{c}{\textbf{GitHub issue from maui (Driver)~\citep{maui-issue}}} \\\hline
Title: &
  SwedishStemmer (and DutchStemmer?) not thread safe \#10 \\\hline
Cmnt \#1: &
  \begin{tabular}[c]{@{}l@{}}While using a Swedish language Maui Server project concurrently from multiple \\ processes, I got several 500 Internal Server Errors with the following traceback:\\\textit{...\textbf{AnalysisEngineProcessException}: Annotator processing failed.}\\ \textit{...Caused by: java.lang.\textbf{StringIndexOutOfBoundsException}: String index out of range: 8}\\ \\ The root cause seems to be that \textbf{the Snowball stemmer used by SwedishStemmer is}\\ \textbf{not thread safe}. (see a \textcolor{red}{\underline{similar issue}} in another project)...\end{tabular} \\ \hline
Cmnt \#2: &
 \begin{tabular}[c]{@{}l@{}}\texttt{osma:} Make SwedishStemmer and DutchStemmer thread safe. Adds tests. Fixes \#10 \end{tabular}   
\end{tabular}

}


\vspace{5mm}

\resizebox{\linewidth}{!} {
     \begin{tabular}{l|l}
\multicolumn{1}{c}{} & \multicolumn{1}{c}{\textbf{GitHub issue from deeplearning4j (Navigator)~\citep{deep-issue}}}\\\hline
Title: &
  Stemmer exception when training word2vec with the supplied tweets\_clean.txt file! \#31\\\hline
Cmnt \#1: &
  \begin{tabular}[c]{@{}l@{}}Hi, I've tried to run the word2vec (using the supplied Uima tokenizer)  and I keep\\ getting this error for many of the words in the sentences...
\\ \\ \textit{...\textbf{AnalysisEngineProcessException}: Annotator processing failed.}\\ \textit{...Caused by: java.lang.\textbf{StringIndexOutOfBoundsException}: String index out of range: 7}\end{tabular} \\ \hline
Cmnt \#2: &
  \begin{tabular}[c]{@{}l@{}}It seems that the reason is that the \textbf{SnowballStemmer IS NOT thread safe}...
\end{tabular}
\end{tabular}
}

\end{table*}

While the concept of ``similar bugs'' could offer an alternative solution for the test generation problem, the prevalence of ``similar bugs'' across different projects and their common characteristics have not been explored. In fact, for open-source projects, software developers have been using the concept of ``similar bugs'' for fixing bugs by referring to a related GitHub issue. Table~\ref{tab:similar-bug} shows a real world example of similar bugs where the \texttt{Maui}'s developer refer to the related issue in \texttt{deeplearning4j} (via the ``similar issue'' link). Comparing the two issues, we can observe that they share some similarities, including: (1) throw the same exception (\texttt{AnalysisEngineProcessException} and \texttt{StringIndexOutOfBoundsException}), (2) depend on the same \texttt{Snowball} library\footnote{Snowball library contains a set of stemming algorithms} where the \texttt{SnowballStemmer} class is invoked, and (3) have the same type of defect (not thread safe). This example shows the scenario target in this paper: \emph{Given an open GitHub issue $d$, can we find a similar bug from another issue $n$ that have been resolved/closed?} Specifically, we adopt the same metaphor used in pair programming and in collaborative bug finding~\citep{collabbug} where the \emph{driver} writes the code, and the \emph{navigator} reviews the code. In our scenario, the open issue $d$ serves as the driver as it tries to resolve the issue, whereas the previously resolved issue $n$ plays the role of the navigator to provide patch suggestion and explanation. By using $n$ as navigator, we foresee several potential benefits: (1) the bug fixing commit in $n$ could contain fixes that require multi-line edits and are not restricted to a predefined set of bug fix patterns, (2) the comments in $n$ is written in natural language and could provide additional context for understanding the root cause of the bug, and (3) we are not restricted to fixing specific type of defects.


To better understand the characteristics of similar bugs across different projects, we conducted an empirical study on \numstudy{} pairs of real GitHub issues collected from popular open-source Java projects. The study aims to explore the following three research questions:
\begin{description}[leftmargin=*]
        \item[RQ1:] \emph{What are the common characteristics that define similar bugs?} 
         \item[RQ2:] \emph{Could similar bugs help developers in testing, debugging or fixing?} 
         \item[RQ3:] \emph{What strategies can we use to detect similar issues?} 
         \end{description}


By investigating these research questions, we have derived several interesting findings. Specifically, we found that similar bugs include bugs that occur when two GitHub projects share the (1) same libraries, (2) same functionality, (3) same steps to reproduce, (4) same configuration/environment, (5) same outcome, or (6) same error/exception. Our study also shows that similar bugs could help in providing more context in resolving open issues (\studycontext{} of the studied GitHub issues), and in fixing the bug (\studyrepair{} of the studied issues). These results give promising evidence about the usefulness of similar bugs in helping software developers in debugging and bug fixing. 


Inspired by the results of our study, we designed \mytool{} based on two important insights. Our key algorithmic insight is to represent each characteristic that defines the similarity between the driver and the navigator as pluggable analysis, and to combine multiple similarity analyses to select the best navigator. By designing each similarity metric as pluggable analysis, we could easily extend \mytool{} to support different types of programs, including Java programs and Android apps evaluated in this paper. Our performance insight is search for relevant GitHub issues via two phrases: (1) online query generation that searches for relevant GitHub issues using off-the-shelf GitHub API, and (2) offline query generation that re-ranks the GitHub issues based on multiple similarity analyses. The online query generation enables us to search broadly for more than millions of issues hosted in GitHub instead of restricting to a small prebuilt database, whereas the offline query generation allows us to deploy more in-depth similarity analyses that require downloading and analyzing source files. 

Overall, our contributions can be summarized as follows:
\begin{description}[leftmargin=*]
\item[Concept.]  We introduce the concept of \cofix, to the best of our knowledge, the first approach that exploits the fact that similar bugs appear even across different software projects to assist developers in debugging and repair. We reformulate the automated patch generation problem as a similar bug recommendation problem where the similar bug may contain a patch and explanation for the root cause. 
\item[Study.] Prior research on duplicate bug report detection rely on a narrow definition of \emph{similar bugs} (bugs that involve handling at least 50\% of common files)~\citep{yang2016combining,rocha2016empirical,rocha2015nextbug}, whereas similar bugs are used to discover new bugs in prior work on \cofind~\citep{collabbug}. To the best of our knowledge, we present the first comprehensive study on the characteristics and usefulness of similar bugs in GitHub.
\item[Technique.] We introduce a repair technique that combines \faultloc{} and a set of pluggable offline similarity analyses for selecting a GitHub issue with the most similar bug to the given open issue. Our \faultloc{} automatically generates query to search online for more than millions of GitHub issues, whereas our similarity analyses exploit information on code changes, library dependencies, permission sets, and UI components in Android apps to select the relevant GitHub issues.
\item[System.] We propose and implement \mytool{}, a new bug report recommendation system for automating \cofix. Given an GitHub issue $d$, \mytool{} will automatically select the best GitHub issue which can be used for debugging and fixing the bug in $d$. The source code for \mytool{} and our experimental data are publicly available at \url{https://crossfix.github.io/}~\citep{our-web}.
\item[Evaluation.] We evaluate the effectiveness of \cofix{} on \numeval{} open GitHub issues (\numevaljava{} issues from Java projects and \numevaland{} issues from Android projects). Our evaluation shows that the issues recommended by \mytool{} could help developers in debugging and fixing the bugs in the open GitHub issues. We have reported similar bugs for \numcomment{} open issues, where \numreplypos{} of these issues received positive feedback. 

\end{description}

%


\begin{table*}[t]
\centering
 \caption{A similar bug between Typesafe Config and GeoTools}
\label{tab:similar-bug-motivate}
\begin{tabular}{p{0.8\linewidth}}
\multicolumn{1}{c}{\textbf{Open GitHub issue from Typesafe Config (Driver)~\citep{config-issue}}} \\\hline
UTFDataFormatException in SerializedConfigValue.scala:314 \\\hline
\begin{tabular}[c]{@{}l@{}}The exact message is :\\\textit{\scriptsize java.io.\textbf{UTFDataFormatException: encoded string too long}: 93067 bytes}\end{tabular}


\begin{tabular}[c]{@{}l@{}}\textit{\scriptsize at java.io.DataOutputStream.writeUTF(...)}...\end{tabular}


\begin{tabular}[c]{@{}l@{}}\\Java DataOutputStream's writeUTF method states that it cannot write \\ more than 65535 bytes...when serializing a Config object, if any String is \\ more than 65535 bytes, it is not serializable... 
\end{tabular} \\\hline

\end{tabular}

\vspace{5mm}
\begin{tabular}{p{0.82\linewidth}}
     \multicolumn{1}{c}{\textbf{Pull request from GeoTools (Navigator)~\citep{geot-issue}}}\\\hline
     [GEOT-6237] Support for big String (byte length > 65535) ...\\\hline

     \begin{tabular}[c]{@{}l@{}}Writing a String attribute with a big size (String bytes greater than 65535\\ bytes) on SimpleFeatureIO we got an exception:\end{tabular}
     
     \begin{tabular}[c]{@{}l@{}}\textit{\scriptsize java.io.\textbf{UTFDataFormatException: encoded string too long}: 71530 bytes}\end{tabular}
     
     \begin{tabular}[c]{@{}l@{}}\\Seems like due to DataOutputStream.writeUTF String size limit,\\SimpleFeatureIO doesn't allow to write/read big String sizes...\textbf{This PR}\\\textbf{adds support for big String sizes using a flag...}\end{tabular}
     \\\hline
     
\end{tabular}

\end{table*}

\section{Motivating Example}
\label{sec:example}
We demonstrate \mytool{}'s workflow using a similar bug in an evaluated Java project. Table~\ref{tab:similar-bug-motivate} shows the similar bug between from the Typesafe Config (we refer to this library as \emph{Config} for short) and the GeoTools projects. Config~\footnote{https://lightbend.github.io/config/} is a configuration library for JVM languages, whereas GeoTools~\footnote{https://github.com/geotools/geotools} is a library that provides tools for geospatial data. The first row of Table~\ref{tab:similar-bug-motivate} denotes the titles of the two issues~\footnote{as GitHub API basically treats a pull request (PR) as an issue with a patch, we refer to \emph{PR} as a GitHub issue in this example}, and the second row of the table gives their contents. As stated in the second row of the table, the exceptions occur due to the \texttt{String} size limit imposed by the \texttt{DataOutputStream.writeUTF} method. Given the issue $d$ from Config, \mytool{} recommends a relevant issue using the steps below:


\begin{itemize}[leftmargin=*,nosep]
\item For GitHub issues with stack trace, \mytool{} automatically extracts this information from $d$ to generate the query ``UTFDataFormatException encoded string too long in:body,comments''. Note that we added ``in:body,com-ments'' to limit the search for text body and comments (instead of searching in their titles) because stack trace is usually written in an issue's text body. 
\item \mytool{} passes our automatically generated query to GitHub API to search for closed issues in Java projects. Instead of searching through other search engines (like Google), searching through GitHub has several benefits due to its pull-based development nature. As contributors in GitHub usually communicate changes by opening an issue or a pull request (PR) \citep{gousios2016work}, code changes are usually associated with an issue/PR which may have (1) explanation for the context in which the bug occurs (e.g., the ``This PR...'' in Table~\ref{tab:similar-bug-motivate} describes the intention of the patch) and (2) the stack trace information that allows us to search for the corresponding fixes. As \mytool{} essentially performs fault localization by generating queries using stack trace information, we call this step \emph{\faultloc}. For our query, GitHub API returns 10 relevant issues in Java projects, where the GeoTools's PR is ranked fourth in the list. 
\item Given the search results from GitHub API, \mytool{} re-ranks the relevant issues based on our similarity analyses and issue quality ranking. For Java projects, \mytool{} computes code similarity and dependency similarity. For each of returned issue $n$, \mytool{} measures code similarity if $n$ contains a patch. When \mytool{} ranks the GeoTools's issue, since $n$ has a patch, \mytool{} downloads $n$'s code and extracts $patch_{Geo}$. Then, it computes the similarity between the Java files in Config and the modified Java files in $patch_{Geo}$. 
As Config does not have any dependencies, we do not calculate its dependency similarity.
\item In the original GitHub's returned results, other issues either have no fix or have fixes that do not share similar code with Config. \mytool{} selects the GeoTools's issue because (1) it has the highest code similarity (59.4\%) among all relevant issues, and (2) it is ranked high by our issue quality ranking function as it contains a fix and has reasonable amount of content (number of words). The output of \mytool{} is a sorted list of issues with the GeoTools's issue as top 1.
\end{itemize}

\vspace{3mm}

\begin{lstlisting}[language=java,basicstyle=\scriptsize,caption=Manually adapted patch for similar bug in Typesafe Config,upquote=true,aboveskip=0pt,belowskip=0pt,label={lst:adaptedpatch},linebackgroundcolor={%
\ifnum\value{lstnumber}=1
            \color{black!20}
\fi
\ifnum\value{lstnumber}=2
            \color{black!20}
\fi
\ifnum\value{lstnumber}=6
            \color{black!20}
\fi
\ifnum\value{lstnumber}=7
            \color{black!20}
\fi
\ifnum\value{lstnumber}=8
            \color{black!20}
\fi
\ifnum\value{lstnumber}=9
            \color{black!20}
\fi
\ifnum\value{lstnumber}=12
            \color{black!20}
\fi
\ifnum\value{lstnumber}=18
            \color{black!20}
\fi
},frame=single]
//added a new method split
^$\Plus$^ private static Collection<String> split(...) {...}
@@ -311,7 +321,19 @@ private static void writeValueData(...)
^$\Minus$^  out.writeUTF(((ConfigString) value).unwrapped());
^$\Plus$^  String strVal = ((ConfigString) value).unwrapped();
^$\Plus$^  List<String> values = new ArrayList<>();
^$\Plus$^  if (strVal.getBytes().length >= MAX_BYTES_LENGTH) {
^$\Plus$^      values.addAll(split(strVal, MAX_BYTES_LENGTH/2));
^$\Plus$^  } else { values.add(strVal);}
...
^$\Plus$^  out.writeInt(values.size());
^$\Plus$^  for (String evalue : values) {
^$\Plus$^      out.writeUTF(evalue);  }
@@ -355,7 +380,12 @@ private static AbstractConfigValue readValueData(...)
^$\Minus$^  return new ConfigString.Quoted(origin,in.readUTF());
^$\Plus$^  int readTimes = byteLength/MAX_BYTES_LENGTH + ((byteLength%MAX_BYTES_LENGTH==0)?0:1);
^$\Plus$^  StringBuilder sb = new StringBuilder();
^$\Plus$^  for (int i = 0; i < readTimes; i++){
^$\Plus$^      sb.append(in.readUTF());  }
^$\Plus$^  return new ConfigString.Quoted(origin,sb.toString());\end{lstlisting}

\vspace{3mm}

\begin{lstlisting}[language=java,basicstyle=\scriptsize,caption=Original patch for GeoTools,upquote=true,aboveskip=0pt,belowskip=0pt,label={lst:geopatch},linebackgroundcolor={%
     \ifnum\value{lstnumber}=1
                 \color{black!20}
     \fi
     \ifnum\value{lstnumber}=2
                 \color{black!20}
     \fi
     \ifnum\value{lstnumber}=10
                 \color{black!20}
     \fi
     \ifnum\value{lstnumber}=7
                 \color{black!20}
     \fi
     \ifnum\value{lstnumber}=8
                 \color{black!20}
     \fi
     \ifnum\value{lstnumber}=9
                 \color{black!20}
     \fi
     \ifnum\value{lstnumber}=13
                 \color{black!20}
     \fi
     \ifnum\value{lstnumber}=23
                 \color{black!20}
     \fi
     },frame=single]
     //added a new method split
     ^$\Plus$^ private static Collection<String> split(...) {...}
     @@ -104,7 +109,26 @@ void writeAttribute(AttributeDescriptor ad, Object value) throws IOException {
     ^$\Minus$^  raf.writeUTF((String) value);
     ^$\Plus$^  if (isBigString(ad)) {
     ^$\Plus$^    String strVal = (String) value;
     ^$\Plus$^    List<String> values = new ArrayList<>();
     ^$\Plus$^    if (strVal.getBytes().length >= MAX_BYTES_LENGTH) {
     ^$\Plus$^       values.addAll(split(strVal, 32767));
     ^$\Plus$^    } else { values.add(strVal);}
     ...
     ^$\Plus$^       raf.writeInt(values.size());
     ^$\Plus$^       for (String evalue : values){
     ^$\Plus$^         raf.writeUTF(evalue);     }
     ^$\Plus$^  } else {// normal string encoding
     ^$\Plus$^     raf.writeUTF((String) value);
     ^$\Plus$^  }
     @@ -179,7 +210,18 @@ Object readAttribute(AttributeDescriptor ad) throws IOException {
     ^$\Minus$^  return raf.readUTF();
     ^$\Plus$^  if (isBigString(ad)) {
     ^$\Plus$^        int parts = raf.readInt();
     ^$\Plus$^        StringBuilder sb = new StringBuilder();
     ^$\Plus$^        for (int i = 0; i < parts; i++) {
     ^$\Plus$^            sb.append(raf.readUTF());   }
     ^$\Plus$^        return sb.toString(); } ...
\end{lstlisting}

To assess the usefulness of the recommended patch, we manually adapted the patch from the GeoTools's issue ($patch_{Geo}$). The third row of Table~\ref{tab:similar-bug-motivate} shows our adapted patch for the Config' issue and the original patch in GeoTools. By referring to $patch_{Geo}$, we can fix the bug in the Config's issue by making minimal modifications to $patch_{Geo}$. In fact, we can reuse almost all the code in $patch_{Geo}$ in constructing the patch for Config (the highlighted code in Listing~\ref{lst:adaptedpatch} and Listing~\ref{lst:geopatch} is the same). We manually constructed a test case and verified that the adapted patch for Config have successfully fixed the exception. Moreover, we have filed a PR with the patch and it is currently under review. This example shows that \emph{\mytool{} could recommend GitHub issues that may contain complex patches that need multiple insertions and modifications of code}. As the recommended patch involves insertions of an entire method and modifications of multiple lines, existing repair approaches will not be able to synthesize this patch from scratch. Although we need to manually adapted the patch currently, an interesting future direction would be to automatically transplant the recommended patches~\citep{barr2015automated}. 

Moreover, since \mytool{} recommends GitHub issues, it naturally leverages GitHub's infrastructure to associate a patch with the corresponding GitHub issue. As shown in the second row of Table~\ref{tab:similar-bug-motivate}, the two GitHub issues contain detailed explanation for the root cause of the exception. This example illustrates that \emph{the explanation included in the GitHub issues could serve as patch explanation} to help developers in understanding how to properly fix the exception~\citep{monperrus2019explainable,explainpatch}.

\section{A study of similar bugs across different projects in GitHub}
\label{sec:study-cause}
Given an open and unresolved GitHub issues $d$, identifying a relevant GitHub issues $n$ that shares similar bug with $d$ could be useful for localizing and fixing the bug in $n$. To design more effective tools to search for relevant GitHub issues, we need to first thoroughly understand the characteristics of similar bugs, including answering the research questions (\textbf{RQ1}-\textbf{RQ3}) listed in Section~\ref{sec:intro}. We investigate $RQ1$ because prior study shows that similar bugs could exist even across two different applications~\citep{collabbug} but the characteristics of ``similar bugs'' have not been studied. We design $RQ2$ to investigate the usefulness of referring to GitHub issues with similar bugs. We study $RQ3$ to design better tools to better utilize and extract information from GitHub issues with similar bugs. 




To study the relationship between GitHub issues that share similar bug, we manually inspected related GitHub issues. Our goal is to identify the information that we can obtain from GitHub issues with similar bugs. We first obtained a list of GitHub issues by searching for the keywords ``similar bug'' and ``similar problem'' written in Java using \texttt{PyGitHub} (a Python library that interacts with GitHub API v3)~\footnote{https://github.com/PyGithub/PyGithub}. Our study focuses on Java projects because (1) most of the recent automated program repair techniques have been developed for fixing Java programs~\citep{xiong2017precise,jiang2018shaping,mei-et-al-ASE15,wen2018context,liu2019tbar} (including the recent bug report driven repair approach~\citep{bugreportrepair}), which indicates the importance of fixing bugs for Java programs; and (2) prior work on \cofind{} shows the existence of similar bugs across different Android apps while most Android apps are written in Java~\citep{collabbug}. For each issue $d$, our crawler searched for the first corresponding issue $n$ that satisfies two criteria: (1) $d$ mentioned that $d$ and $n$ share similar bugs; and (2) $d$ and $n$ are in different open-source projects ($proj_d$ != $proj_n$, where $proj_i$ indicates the open-source project in which the GitHub issue $i$ was reported). The final output of our crawler is a list of GitHub issues ($d$, $n$) where $d$ and $n$ share similar bugs across different open-source projects. Overall, our crawler searched through 2000 GitHub issues (GitHub API limits each search to 1000 search results~\footnote{https://docs.github.com/en/rest/reference/search}, so we search for the two keywords separately), and identified 385 issues with links to similar bugs. From these 385 GitHub issues, we manually excluded 194 invalid issues (e.g., duplicated issues or irrelevant issues). After filtering the irrelevant issues, we identified \numstudy{} ($d$, $n$) pairs of GitHub issues where $n$ spans across \numstudyproj{} different Java projects. This relatively large number of open-source projects with GitHub issues mentioning similar bugs confirm with our hypothesis that \emph{software developers tend to leverage the information about similar bugs during the discussion of a GitHub issue}.



\subsection{RQ1: Characteristic of similar bugs}

\begin{table}[t]
\centering
 \caption{Common characteristics of similar bugs, and strategies to identify them (\emph{one issue could have multiple characteristics})}
\label{tab:study-tab}
 \resizebox{\linewidth}{!} {
\setlength\tabcolsep{0.9pt}
\begin{tabular}{l|l|r}
Strategy  & Characteristic of similar bugs & \# of Issues\\ \hline
Dependency  & Same library                  & 76 \\ \hline
Code & Same functionality (forks, different language)                                  & 5 \\ \hline
UI            & Same steps to reproduce         & 2 \\ \hline
Permission          & \begin{tabular}[c]{@{}l@{}}Same configuration/environment\\(Permission set, Android/Window version)\end{tabular}  & 6 \\ \hline
Condition   & Same outcome (non-crashing)         & 35 \\ \hline
Stack trace & Same error/exception            & 39  \\ \hline 
\end{tabular}
}
\end{table}

For RQ1, we obtain the characteristic of similar bugs by reading carefully the GitHub issues, focusing particularly on comments made by developers when referring to the similar bugs (i.e., developers usually include explanation why the two bugs are similar). Then, we count the number of GitHub issues that exhibit the identified characteristics. 

The second column of Table~\ref{tab:study-tab} shows the common characteristics that define similar bugs in the evaluated pairs of GitHub issues, whereas the ``\# of Issues'' column denotes the number of GitHub issues that exhibit these characteristics. Our study shows that similar bugs can be defined by bugs that occur when two GitHub projects share the (1) same libraries, (2) same functionality, (3) same steps to reproduce, (4) same configuration/environment, (5) same outcome (exclude crash), or (6) same error/exception. As a pair of GitHub issues ($d$, $n$) can exhibit several similarities (e.g., they could share same libraries and produce the same error as the example in Table~\ref{tab:similar-bug}), we include all similarities in Table~\ref{tab:study-tab} (the category is mutually exclusive). Overall, our study shows that the most common characteristics of similar bugs are:
\begin{itemize}
\item Bugs that share the same library (76/\numstudy{} = 50\%)  
\item Bugs that share the same error/exception (39/\numstudy{} = 26\%)
\item Bugs that share the same outcome (35/\numstudy{} = 23.33\%) 
\end{itemize}

The relatively high percentage of bugs that share the same library \emph{motivates the needs for automated dependency update tools}~\citep{dependabot,greenkeeper,macho2018automatically}.

\subsection{RQ2: Usefulness of similar bugs}
\label{sec:usefulsim}
We investigate the usefulness of similar bugs by checking whether the \numstudy{} issues help developers to perform specific maintenance tasks by scoring them in the following order: (1) repair (most useful), (2) fault localization, (3) finding context of a bug, and (4) closing the issue faster (least useful). Specifically, if the similar bug could help developer directly in locating and fixing the bug, we select the most useful task (``repair'') rather than ``fault localization''. During our analysis, we exclude 64 GitHub issues from \numstudy{} because we could not determine their usefulness as there are no direct reply or discussion about the similar issue. Specifically, given a pair of GitHub issues ($d$, $n$), we assess $n$'s usefulness based on the following descending order (we select the highest usefulness score for $n$ if $n$ can help in several maintenance tasks):
\begin{description}[leftmargin=*]
\item[Repair:] We consider $n$ as helping developer in $d$ to repair the bug in $d$ if: (1) the fix of $n$ could be directly applied to $d$, (2) the fix of $n$ only requires minor modifications before applying to $d$, or (3) the discussion of $d$ mentioned that they could find the fix for $d$ by referring to the information in $n$.
\item[Fault localization:] We consider $n$ as helping developer in $d$ to locate the bug if: (1) $n$ causes the bug in $d$, (2) $n$ includes fault localization information, or (3) the discussion of $d$ mentioned that they could identify the faulty location using the information in $n$. 
\item[Context:] We consider $n$ as helping developer in $d$ to finding more context for the bug in $d$ if the discussion in $d$ explicitly mentioned that they have learned something from $n$ (e.g., they gained some clues after reading $n$) but there is no further discussion about whether the knowledge gained had helped them in resolving the problem or identifying the fault. 
\item[Closing Issue faster:] We consider $n$ as helping developer in $d$ to closing the issue faster if: (1) after referring to the similar issue in $n$, the participants in the discussion of $d$ think that they cannot or do not need to fix the bug and decided to close the issue, or (2) $n$ only provides temporary workaround.
\end{description}

\begin{figure}
    \begin{minipage}[t]{\linewidth}
        \centering
        \includegraphics[height=0.5\linewidth]{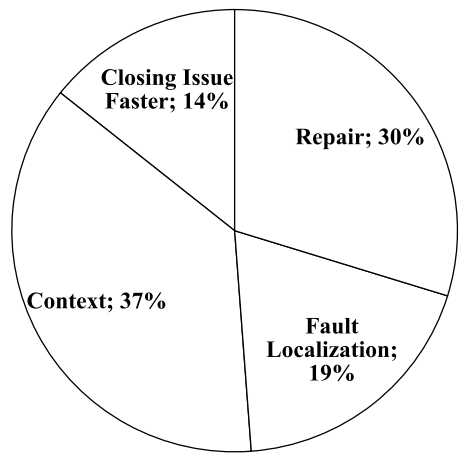}
    \caption{The usefulness of similar bugs among 86 study issues}
    \label{fig:usefulness}
    \end{minipage}
\end{figure}

Figure~\ref{fig:usefulness} shows the usefulness of similar bugs among the 86 studied issues. Our study shows that similar bugs is most helpful in terms of helping developers to find more context (provide hints) for resolving the bug (\studycontext{} of studied GitHub issues). Meanwhile, \studyrepair{} of studied issues show the usefulness of similar bugs in helping developers to fix the bug. Overall, our results shows that \emph{similar bugs could help developers in resolving the GitHub issue in hand, especially in providing more context of the bug and fixing the bug}.





\subsection{RQ3: Strategies to identify similar bugs}
\label{subsec:rqtwo}
After identifying the common characteristics of similar bugs, we need to design strategies to automatically detect these characteristics. Specifically, we need to check if we can detect each of these characteristics from several available sources, including: (1) the description of a GitHub issue, and (2) information within different types of files that can be exploited for detecting these characteristics (e.g., Java projects typically has build files and Java class files, whereas Android project may have additional XML files for declaring app components and permissions). 

Given a pair of GitHub issues ($d$, $n$) that share a similar bug $b$, we derive \numop{} strategies to identify similar bugs based on the rules explained below:

\begin{description}[leftmargin=*]
\item[Dependency:] We check if $d$ and $n$ share the same libraries by comparing the set of dependencies shared by the GitHub issues in $d$ and $n$. For example, in Table~\ref{tab:similar-bug}, both \texttt{Maui} and \texttt{deeplearning4j} share similar dependency (\texttt{Stemmer}). Java dependencies are usually declared in build files. 
\item[Code:] We check if $d$ and $n$ share similar functionalities by comparing the Java code in the repository for $d$ and the code modifications in the patch mentioned in $n$. Ideally, the code similarity checking will be more precise if we can compare the buggy code in $d$ with the patch mentioned in $n$. However, as the information about the buggy code is often missing, we can only check whether the code within the patch for $n$ matches with any code within the entire repository for $d$. 

\item[UI (Android-specific):] We check if $d$ and $n$ share the same steps to reproduce the bug $b$ by comparing their interfaces that allow users to invoke these steps. In Android apps, interfaces refer to UI elements (e.g., buttons and text boxes) that are typically declared in XML files.
\item[Permission (Android-specific):] We check if $d$ and $n$ share similar environments by comparing the set of permissions that $d$ and $n$ acquire. We design this strategy as all Android apps need to declare the set of permissions in their manifest files but there is no such formal requirement for Java projects. The set of permissions is typically declared in XML files. 
\item[Condition:] We check if $d$ and $n$ share the same outcomes by comparing the conditions when the bug $b$ is triggered. This information is usually available in an issue's description.  
\item[Stack trace:] We check if $d$ and $n$ share the same error or exception by comparing their stack traces. For example, in Table~\ref{tab:similar-bug}, the issues in \texttt{Maui} and \texttt{deeplearning4j} share similar stack trace with similar exception. This information is usually available in the description of a GitHub issue.   
\end{description}

\section{Methodology}

\begin{figure*}[t!]
\begin{center}
  \includegraphics[width=\linewidth]{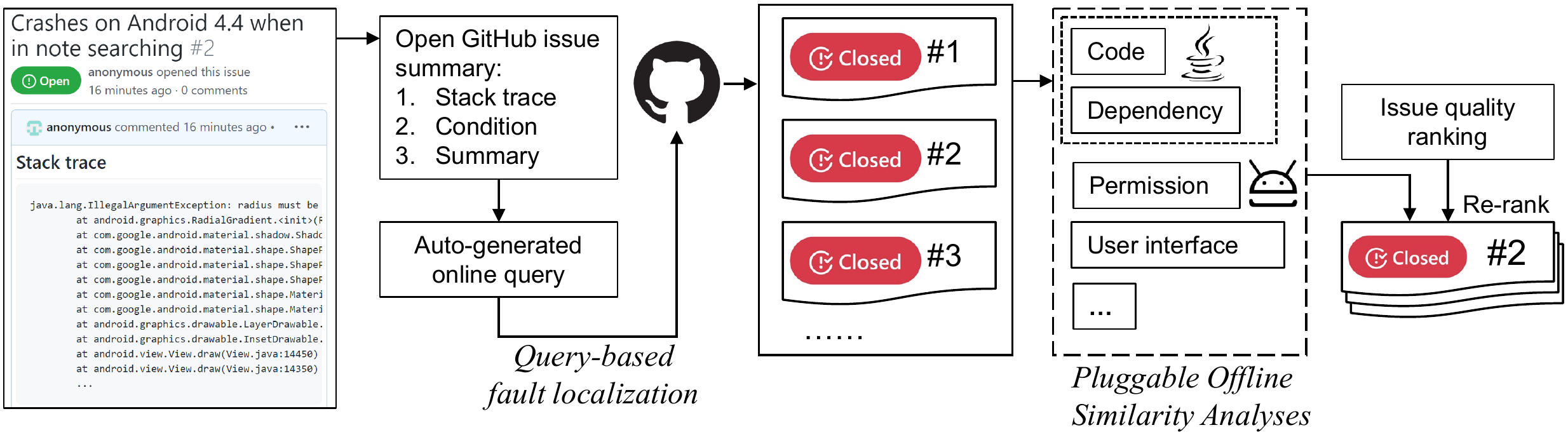}
  \caption{\mytool's recommendation system}
  \label{fig:workflow}
\end{center}
\end{figure*}

Figure~\ref{fig:workflow} presents the overall workflow of \mytool{}. \mytool{} consists of several components: lightweight online query generator, pluggable offline similarity analysis, and issue quality ranking. \mytool{} uses a two-phase approach. In the first phase, given an open GitHub issue $d$, \mytool{} automatically extracts information from $d$ to build a query $q_{online}$ to search for closed GitHub issues with similar (1) exception information, (2) condition that triggers the bug, or (3) general description that summarizes the bug extracted from $d$'s title. In the second phase, \mytool{} goes through the search results for $q_{online}$ and re-rank the list of GitHub issues $n_1, n_2, ..., n_i$. For each issue $n_i$, \mytool{} computes (1) the quality of $n_i$, and (2) the similarity between $d$ and $n_i$ based on multiple analyses. The final output of \mytool{} is a ranked list of GitHub issues. We consider only the top-1 GitHub issue as the best \nav{} that will be used as ``similar bug''. 

\subsection{\faultloccap{}}
\label{subsec:onlinequery}

Existing techniques in automated program repair typically treat test cases as blackbox~\citep{angelix,prophet,martinez2016automatic,genprog-icse09} and tend to ignore the target defect classes that these techniques aim to address~\citep{critical-review}. To find the faulty location that causes the bug, these techniques usually rely on statistical fault localization techniques. Meanwhile, bug report driven repair approaches~\citep{r2fix,bugreportrepair} use information from bug report from the buggy program to either determine the bug type~\citep{r2fix} or perform fault localization~\citep{bugreportrepair}. Different from these techniques, \mytool{} performs \emph{\faultloc} by generating search queries using information extracted from GitHub issues of the buggy program. This new way of fault localization introduces new challenges: (1) the input GitHub issue is written in natural language which is difficult to parse correctly, (2) we need to automatically determine what kind of information in the given GitHub issue that is useful for representing the bug; and (3) the generated query need to be short and precise because GitHub imposes restriction on the length of the search query: queries longer than 256 characters are not supported~\footnote{https://docs.github.com/en/github/searching-for-information-on-github/troubleshooting-search-queries}. 

Given an open issue, \mytool{} first parses its title and its description. As the GitHub issue is written in natural language, \mytool{} pre-processes the natural language text using the same procedure in prior work~\citep{collabbug}. Specifically, we perform tokenization, stopword removal (via Python NLTK library~\citep{nltk}), stemming, and lemmatization. We also exclude the project name (e.g., \texttt{deeplearning4j}) from the query to remove project specification information.

\begin{table*}[t]
  \centering
  \caption{Strategies used in extracting information to build the online query.}
  \label{tab:online-query}
  \setlength\tabcolsep{1pt}
  \begin{tabular}{p{1.5cm}<{\centering}|p{2.5cm}|p{1cm}<{\centering}|p{6cm}}
    \multicolumn{1}{c}{Strategy} & \multicolumn{1}{c}{Description} & \multicolumn{1}{c}{Source} & \multicolumn{1}{c}{Example Text (Upper)/Generated Query (Lower)} \\ 
    \hline
    \multirow{2}{*}{Stack trace} & \multirow{2}{*}{\begin{tabular}[c]{@{}l@{}} Exception/Error \\ type and excep-\\tion/error message\\ in the stack trace \end{tabular}} & \multirow{2}{*}{\begin{tabular}[c]{@{}c@{}}issue\\body\end{tabular}} & \begin{tabular}[c]{@{}l@{}}java.lang.\textbf{NullPointerException}:~\textbf{Attempt}\\\textbf{to invoke virtual method}~'\textbf{java}.\textbf{lang}.\textbf{Ob-}\\\textbf{ject}.\textbf{android}.\textbf{widget}.\textbf{FrameLayout}.\textbf{get-}\\\textbf{Tag(int)}'~\textbf{on a null object reference}\end{tabular} \\ 
    \cline{4-4}
      &  &  & \begin{tabular}[c]{@{}l@{}}NullPointerException attempt to invoke\\virtual method java lang Object android\\ widget FrameLayout getTag int on a null\\object reference \end{tabular} \\ 
    \hline
    \multirow{2}{*}{Condition} & \multirow{2}{*}{\begin{tabular}[c]{@{}l@{}} Condition where \\ the buggy behav-\\ior is triggered \end{tabular}} & \multirow{2}{*}{\begin{tabular}[c]{@{}c@{}} issue\\title \end{tabular}} & \begin{tabular}[c]{@{}l@{}}Stemmer exception when~\textbf{training word2vec}\\\textbf{with the supplied tweets\_lean.txt~file}\end{tabular} \\ 
    \cline{4-4}
      &  &  & \begin{tabular}[c]{@{}l@{}}training word2vec with the supplied \\tweets\_lean.txt file \end{tabular} \\ 
    \hline
    \multirow{2}{*}{Summary} & \multirow{2}{*}{\begin{tabular}[c]{@{}l@{}} summary of the \\bug from title \end{tabular}} & \multirow{2}{*}{\begin{tabular}[c]{@{}c@{}} issue \\title \end{tabular}} & \begin{tabular}[c]{@{}l@{}}\textbf{SwedishStemmer}~(and~\textbf{DutchStemmer}?)~\\not~\textbf{thread safe}~\end{tabular} \\ 
    \cline{4-4}
      &  &  & SwedishStemmer DutchStemmer thread safe
  \end{tabular}
\end{table*}

After pre-processing, \mytool{} automatically extracts information that can be used to build a query for searching through GitHub. To determine the key information that should be included in the generated query for representing the bug, we design our query based on our study (described in Section~\ref{sec:study-cause}), and select only the information that can be directly extracted from the description of a GitHub issues (including stack trace and condition).  
Table~\ref{tab:online-query} shows the strategies that we used to build the online query. As shown in the ``Strategy'' column in Table~\ref{tab:online-query}, \mytool{} generates query based on three strategies: (1) stack trace (highest priority), (2) condition, and (3) bug summary (this strategy is used only as a fallback plan when we cannot apply other strategies). Algorithm~\ref{alg:query} shows our query construction algorithm. Given an issue $i$ with title $title$ and issue body $body$, together with a threshold $n$ (empirically set to 5 to indicate that a query returns too few search results), our approach returns a ranked list of search results. 

\SetKwInput{KwInput}{Input}                
\SetKwInput{KwOutput}{Output}              

\begin{algorithm}[t]
  
  \KwInput{A GitHub issue $i$ with title $title$ and issue body $body$}
  \KwOutput{GitHub Search Results $results$}
  \KwData{Threshold $n$}
  $query$ := ""\;
  $results$ := []\;
  \If{$body$.match("[a-zA-Z.]*(Exception|Error)")}
    {
     \tcc{extract exception type and message}
        $query$ := extractException($body$) + "in:body"\;  
        $results$ : = getGitHubResults($query$)\;
    }
    \If{$len(results)$ < $n$}{ \If{$title$.match("(if|when|while).\textasteriskcentered")}
    {
    \tcc{extract the phrase after if|when|while}
    $query$ := extractCondition($title$) + "in:title"\; 
    $results$ : = getGitHubResults($query$)\;
    }
    
    \Else
    {
        \tcc{remove common words (e.g, a, the)}
    	$query$ := removeStopwords($title$) + "in:title"\;  
    	$results$ : = getGitHubResults($query$)\;
    	\If{$len(results)$ <  $n$}{
    	$query$ := removeStopwords($title$)\;
    	$results$ : = getGitHubResults($query$)\;
    	}
    }
    }
    
\Return{$results$}
\caption{Our GitHub query construction algorithm}
\label{alg:query}
\end{algorithm}

We describe each strategy in detail below:

\noindent\textbf{Stack trace.} Our \faultloc{} uses stack trace information to represent the bug because (1) this information is often included in an issue's text body, (2) stack trace may contain the core elements of a patch explanation, including the cause (the root cause exception/error), and the consequence (crash)~\citep{explainpatch}, and (3) prior study has shown the effectiveness of using stack trace to locate~\citep{sinha2009fault,gu2019does,wu2018changelocator,han2012performance}, and to repair runtime exceptions~\citep{tan2018repairing}. The key differences between our approach and existing fault localization techniques are: (1) our stack trace information is embedded in an issue's text body, and (2) our stack trace may be incomplete because some users may include only partial stack trace in the GitHub issue. Given a GitHub issue $d$, our stack trace parser extracts $d$'s text body and searches for stack trace information. Specifically, our stack trace parser identifies stack trace by searching for the regular expression ``[a-zA-Z.]*(Exception|Error)'' (line 3 in Algorithm~\ref{alg:query}). Then, our parser searches for the ``Caused by'' keyword to locate the root cause of the crash. If the parser fails to get the root cause due to incomplete stack trace, it then searches for the first Exception/Error in the first line of the extracted stack trace. This strategy outputs a query that contains the name and the message of the thrown Exception/Error in the given GitHub issue. We give an example to explain the input (``Example Text'') and output (``Generated Query'') of this step in the first row of Table~\ref{tab:online-query}.

\noindent\textbf{Condition.} As the condition under which a bug is triggered is a core element of a patch explanation~\citep{explainpatch}, \mytool{} searches for the phrases that represent a condition in the issue's title. Specifically, it looks for the expression \texttt{$(if|when|while).*$} (line 7 in Algorithm~\ref{alg:query}). For example, the phrase \emph{training word2vec with the supplied tweets\_clean.txt file} in the Maui's issue in Table~\ref{tab:similar-bug}, and second row of Table~\ref{tab:online-query}. 

\noindent\textbf{Bug summary.} Although more advanced text summarization techniques could be used to generate a precise summary, we choose to use the text extracted from the issue title because this information is directly available and it provides a fast way to obtain an overall summary of the issue. As shown in the last row in Table~\ref{tab:online-query}, we build the query by removing symbols and stopwords (``and'' and ``not'') from the issue title (line 11--15 in Algorithm~\ref{alg:query}).




After generating the online query via our \faultloc{}, we passed the generated query to GitHub API to search for relevant GitHub issues. 

\subsection{Pluggable Offline Similarity Analyses}
\label{subsec:similarity}

To further rank the relevant GitHub issues based on the initial search results returned by the online query, we design the analysis based on the results discussed in Section~\ref{subsec:rqtwo}. To diagnose different types of defects, \mytool{} applies multiple \emph{pluggable similarity analyses} for checking different types of source files with each analysis applied when it is required. This design allows us to analyze both Java projects and Android apps by reusing several common similarity analyses and applying specialized analyses to Android apps. Specifically, for Java and Android projects, we perform (1) code similarity analysis, and (2) dependency similarity analysis. For Android projects, we additionally perform (1) UI similarity analysis, and (2) permission set similarity analysis. 

In general, given two GitHub issues ($d$, $n$), each analysis for a given characteristic $c$ checks if (1) $c$ is mentioned in both issues $d$ and $n$, and (2) $c$ defines the similarity between $d$ and $n$. We describe the analysis for each characteristic as below:

 \noindent\textbf{Code Similarity.} Given two GitHub issues ($d$, $n$) where $n$ contains a patch $p_n$ which fixes the bug, we measure code similarity between $p_n$ and $d$ using the equation below:
\begin{equation}
   CodeSim(n, d, p_{n})=sim(\forall f \in repo_{d}, \forall f \in p_n)
\end{equation}

Specifically, we compute the code similarity between all files that are modified in $p_n$ and all files in the $d$'s repository. 

\noindent\textbf{Dependency Similarity.} Given two GitHub issues ($d$, $n$) and a function $Dep(i,repo_{i})$ that returns the package dependencies mentioned in a given issue $i$ and declared in the dependency files in $i$'s repository, we measure dependency similarity between $d$ and $n$ using the equation below: 
\begin{equation}
        DepSim(d, n) = Sim(Dep(d, repo_{d}), Dep(n, repo_{n}))
\end{equation}

Particularly, we compute the similarity between the package dependencies mentioned in $d$ and that are mentioned in $n$. 

\noindent\textbf{Permission Similarity.} Given a pair of GitHub issues ($d$, $n$) and a function $Per(i,repo_{i})$ that returns the permissions mentioned in a given issue $i$ and declared in the app manifest files in $i$'s repository, we calculate permission similarity between $d$ and $n$ using the equation below: 
\begin{equation}
        PermSim(d, n) = Sim(Per(d, repo_{d}), Per(n, repo_{n}))
\end{equation}

Specifically, we measure the similarity between the permissions mentioned in $d$ and that are mentioned in $n$. 

\noindent\textbf{UI Similarity.} Given a pair of GitHub issues ($d$, $n$) and a function $UI(i,repo_{i})$ that returns the UI elements mentioned in a given issue $i$ and declared in the XML files in $i$'s repository, we compute UI similarity between $d$ and $n$ using the equation below: 
\begin{equation}
        UISim(d, n) = Sim(UI(d, repo_{d}), UI(n, repo_{n}))
\end{equation}

Particularly, we calculate the similarity between the UI components mentioned in $d$ and that mentioned in $n$. 

\noindent\textbf{Similarity Measure.} For the dependency similarity, the permission similarity and UI similarity, we use \emph{Overlap Coefficient}~\citep{kowalski2010information} to calculate the overlap between the set of keywords mentioned in a GitHub issue and the corpus (words obtained from package dependencies, permission set, and the names of UI elements). Namely, Overlap Coefficient is defined as:


\begin{equation} \label{eq:overlap}
    \operatorname {overlap} (X, Y)={\frac {|X\cap Y|}{\min(|X|, |Y|)}}
\end{equation}

Note that the Overlap Coefficient ranges between $[0, 1]$. If $X$ is the subset of $Y$ or vice versa, then $\operatorname {overlap} (X, Y)$ is equal to 1. We select Overlap Coefficient instead of the Jaccard Similarity \citep{jaccard1901etude} and Dice Similarity \citep{dice1945measures} because (1) it is widely used in prior recommendation systems for bug reports~\citep{rocha2016empirical,collabbug,moin2012assisting}, and (2) it is sensitive to the size of the two sets which is suitable for our task since the search query is usually shorter than the corpus in the database.

\subsection{Selecting the most relevant issue}
\label{subsec:select}

Apart from using multiple similarity analyses, we select the most relevant GitHub issue based on their quality.

\noindent\textbf{Issue quality ranking.} Ideally, the GitHub issues recommended by \mytool{} should contain rich information to assist developers in debugging and fixing. According to prior study, developers expect a good report to contain: (1) \emph{steps to reproduce}, (2) \emph{observed and expected behavior}, and (3) \emph{stack traces}~\citep{zimmermann2010makes}. Hence, we further rank the GitHub issues based on its qualities. Specifically, we use several metrics used in prior work~\citep{collabbug} to evaluate the quality of an issue $n$, including: (1) the number of words in $n$'s text body, (2) if $n$ contains a commit hash (fixed/not fixed), (3) the number of comments that $n$ received, and (4) the number of descriptive keywords that $n$ contains (e.g., ``reproduce'', ``defect'').

\noindent\textbf{Selecting the best GitHub issue.} With multiple similarity analyses (defined in Section~\ref{subsec:similarity}) and the score for the issue quality as factors that affect the ranking of a GitHub issue, we calculate the ranking score $S(n, W)$. Given an issue $n$, we calculate the ranking score $S(n, W)$ as below:

\begin{equation}
    S(n, W)=\sum_{i=1}^{n} f_{i}(n) \times w_{i}
\end{equation}

where $f_{i}(n)$ denotes the value of factor $f_{i}$ on issue $n$; and $w_{i}$ denotes the weight for a factor $f_{i}$. For the factors that affect issue quality ranking, we select the weights used in prior work~\citep{collabbug} ($w_{issue\_length}$ = 0.0714, $w_{num\_comment}$ = 0.1428). For other factors, we perform a grid search to tune their weights using the dataset in Section~\ref{sec:study-cause}. Specifically, we selected $w_{CodeSim}$ = 0.1428, $w_{DepSim}$ = 0.2142, $w_{PermiSim}$ = 0.2142, $w_{UISim}$ = 0.2142.

\subsection{Implementation}
For calculating code similarity, we leverage JPlag~\citep{prechelt2002finding}. JPlag first converts each program into a string of tokens, and then compares the two programs by trying to cover one of the programs with sequences from the other program using the maximum similarity algorithm~\footnote{https://jplag.ipd.kit.edu/example/help-sim-en.html}. While there are many code plagiarism detection engines that can calculate code similarity~\citep{prechelt2002finding,ahtiainen2006plaggie,luo2014semantics,martins2014plagiarism,lancaster2004comparison,schleimer2003winnowing}, we select JPlag because (1) it is open-source and actively maintained, (2) it offers offline similarity analysis, and (3) it has been shown to be one of the most effective code similarity analyzers~\citep{ragkhitwetsagul2018comparison,7522248}. To compute the dependency similarity, we support projects that use Gradle~\footnote{https://gradle.org/} and Maven~\footnote{https://maven.apache.org/} for compilations because (1) they are among the most popular open source build automation tools, and (2) Gradle has been widely used in both Java and Android projects. We compute UI similarity by modifying existing implementation of Bugine, a tool that recommends relevant GitHub issues based on UI similarity~\citep{collabbug}. The main modification to Bugine is to add the support for incorporating the open GitHub issue $d$ into the UI similarity calculation.

\section{Evaluation}
There are several related approaches that we have considered for comparison, including NextBug and GitHub.
Although NextBug~\cite{rocha2015nextbug} also recommends bug reports, it relies on traditional bug tracking systems like Bugzilla instead of GitHub, and it requires bug component to be specified for finding bug reports with similar text descriptions. There are several key differences between Bugzilla and GitHub that makes NextBug not suitable for recommending GitHub issues: (1) a bug report in Bugzilla contains a field for users to specify the bug component~\footnote{https://www.bugzilla.org/docs/2.18/html/bug\_page.html}, whereas such field does not exist in a GitHub issue (users may choose to indicate the bug component by adding a tag but most GitHub issues do not contain any tag); (2) the bugs in Bugzilla are only reported within the same organization (e.g., Mozilla) and the bug reports for each organization are hosted under separate links in Bugzilla, whereas GitHub is a site that hosts a wide variety of projects from many organizations across the world; and (3) GitHub provides hosting for millions of software repositories and includes a version control system, whereas Bugzilla is only a bug tracking system (does not store any source code or commit histories that can be used for finding similar bugs). As NextBug does not support GitHub issues and a fair comparison will not be possible without considering the unique features in GitHub, we did not compare our tool against NextBug. Another related approach is the search feature in GitHub that can be used to narrow down the search for issues that contain certain keywords. We select GitHub as our baseline for comparison since searching through GitHub is the first step of approach.

We perform evaluation on the effectiveness of \mytool{} to address the following research questions:
\begin{description}
\item[Q1] What is the overall performance of \mytool{} in ranking relevant GitHub issues?
\item[Q2] How useful are the \mytool{}'s recommended issues?
\end{description}

\subsection{Experimental Setup}
We evaluate \mytool{} on a total of \numeval{} open GitHub issues. Among these \numeval{} issues, \numevaljava{} issues are from \numevalandproj{} Java projects, whereas \numevaland{} issues are from \numevalandproj{} Android apps.

\noindent\textbf{Selecting open issues.}
To select open GitHub issues, we build a crawler that automatically searches for open issues from popular Java projects and Android apps. We obtained a list of open source Android apps from F-Droid. For Java and Android projects, our crawler selects projects that have (1) the greatest number of stars, (2) recent commits within one year, and (3) no overlap with the projects in our study (Section~\ref{sec:study-cause}). To ensure diversity of the considered projects, our crawler got the first \numiss{} open issues for each project from GitHub that do not have any bug fixing commit. It is challenging to search for open GitHub issues automatically because an issue either (1) has a pull request but the developers did not include its link, (2) has too little information to verify its validity, (3) has irreproducible bug, (4) is not bug-related (question/feature) so we need to manually filter these invalid issues. From our initial list of 192 issues for Java projects and 142 issues for Android apps, we removed 40 invalid issues for Java projects and 45 invalid issues for Android apps. 

Table~\ref{tab:stat} shows the statistics of the evaluated open GitHub issues. Overall, the selected projects are diverse in terms of sizes and functionalities (Java projects include VM, compilers, etc., whereas Android apps include notes app, file sharing app, etc.). Information about the selected projects is available at our website~\citep{our-web}.   




\noindent\textbf{Evaluation metrics.}
For $Q1$, we evaluate the overall ranking performance of \mytool{} by using two measures used in previous evaluations of recommendation systems~\citep{collabbug,Zhou:2012:BFM:2337223.2337226,Ye:2014:LRR:2635868.2635874}:
\begin{description}[leftmargin=*]
\item[Prec@k.] This computes the retrieval precision over the top $k$ documents in the ranked list:
\begin{equation} \label{eq:preck}
\operatorname {Prec@k}={\frac{\#\ of\ relevant\ docs\ in\ top\ k}{k}}
\end{equation}
We measure the precision at k=1, 3, 5. 
\item[Mean Reciprocal Rank (MRR).] For each query $q$, the MRR measures the position $first_q$ of the first relevant document in the ranked list~\citep{voorhees1999trec}: 
\begin{equation} \label{eq:mrr}
\operatorname {MRR}={\frac { 1 } { | Q | } \sum _ { q = 1 } ^ { | Q | } \frac { 1 } { \operatorname { first } _ { q } }}
\end{equation}
A higher MRR value denotes better ranking performance. 

\end{description}

\begin{table}[!t]
\small
\centering
\caption{Statistics of the evaluated Java projects and Android apps}
  \label{tab:stat}
\begin{tabular}{lr|r}
                       & \multicolumn{1}{l|}{Java} & \multicolumn{1}{l}{Android} \\\hline
Total open issues      & 152                      & 97\\ 
Total projects         & \numevaljavaproj{}                      & \numevalandproj\\
KLOCs & 5--2140k & 8--683k\\
\end{tabular}
\end{table}

To assess the usefulness of a recommended issue $n$ in $Q2$, we use the same criteria as $\rqcat$ (we check if $n$ helps in repair, fault localization, context, or closing issue faster). For all relevant issues, we reported them to the developers.


All experiments were conducted on a machine with Intel (R) Core (TM) i7-8700 CPU @3.2 GHz and 32 GB RAM.

\subsection{Q1: Ranking Performance of \mytool} 
\begin{table}[!t]
\caption{Ranking performance of \mytool{}}
  \label{tab:rankperform}
\centering
\begin{tabular}{l|rr|rr}
       & \multicolumn{2}{c|}{Java} & \multicolumn{2}{c}{Android} \\\cline{2-5}
                 & \multicolumn{1}{l}{GitHub} & \multicolumn{1}{l|}{\mytool{}} & \multicolumn{1}{l}{GitHub} & \multicolumn{1}{l}{\mytool{}} \\ \hline
Average Prec@1 & \preckoneavgjavagit & \textbf{\preckoneavgjava{}} & 0.4433 & \preckoneavgand{}          \\
Average Prec@3 & 0.3181 & 0.3181          & 0.4433 & \textbf{0.4467} \\
Average Prec@5 & 0.3163 & 0.3163          & 0.4454 & 0.4454          \\\hline
Average MRR    & 0.1168 & \textbf{0.1401} & 0.0485 & \textbf{0.0619}\\ \hline
\# relevant                      & \multicolumn{2}{r|}{\numjavarel}                       & \multicolumn{2}{r}{\numandrel}                       \\
\end{tabular}
\end{table}

Table~\ref{tab:rankperform} shows \mytool{}'s ranking performance results. As \mytool{} re-ranks the original GitHub returned results using multiple similarity analyses, we compare the performance of the original GitHub's ranking (``GitHub'' column in Table~\ref{tab:rankperform}) versus the re-ranked results by \mytool{} (``\mytool{}'' column in Table~\ref{tab:rankperform}) to assess the effectiveness of our similarity analyses. 

\noindent\textbf{Java versus Android.} Compared to Java projects, the average Prec@$k$ results for Android projects are generally higher. For example, the average Prec@1 is \preckoneavgand{} for Android projects versus \preckoneavgjava{} for Java projects. In contrast, the average MRR values for Java projects are higher than that for Android projects. The higher Prec@$k$ for Android projects is because Android projects have less relevant GitHub issues (25 versus 75 relevant issues) and many queries for Android projects have zero search results (Prec@$k$=1 for zero results).

 
\noindent\textbf{Effectiveness of similarity analyses.} As our approach aims to select the best issue as the similar bug, we use MRR for evaluating the effectiveness of our similarity analyses because MRR gives greater importance to the first relevant item. Comparing the results for the ``GitHub'' column and the ``\mytool{}'' column in Table~\ref{tab:rankperform}, we observe that \emph{our similarity analyses helps to improve MRR values for both Java and Android projects}. Specifically, the average MRR for Java projects increases from 0.1168 to 0.1401 ($\approx 20\%$ increment), whereas the average MRR for Android projects increases from 0.0485 to 0.0619 ($\approx 28\%$ increment).

\begin{framed}
\noindent\textbf{Answer to Q1:} The Prec@k and the MRR values show that \mytool{} is able to recommend relevant issues for both Java and Android projects. 
\end{framed}

\subsection{Q2: Usefulness of the recommended issues} 

 \begin{figure}[!t]
    \begin{minipage}[t]{\linewidth}
        \centering
        \includegraphics[height=0.5\linewidth]{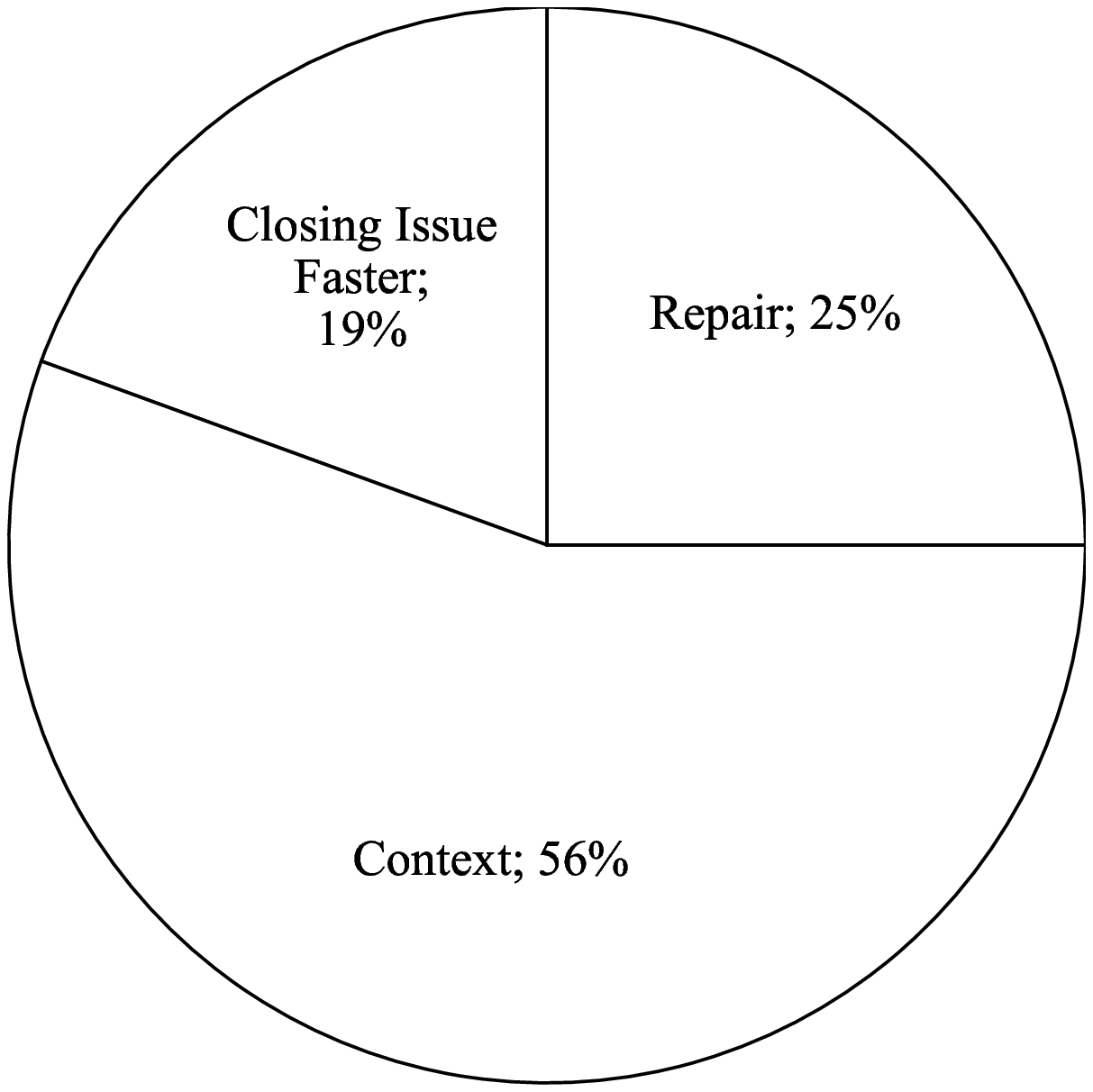}
    \caption{The usefulness of the issues recommended by \mytool{}}
\label{fig:usefulness-mytool}
    \end{minipage}
\end{figure}

Table~\ref{tab:rankperform} shows that \mytool{} recommended \numjavarel{} issues for Java projects and \numandrel{} issues for Android projects. Given a relevant issue $n$, we evaluate Q2 by manually checking if $n$ helps in repair, debugging, finding context about the bug or closing the open issue. In total, there are \numcomment{} open GitHub issues $d$ with at least one relevant recommended issues $n$. Figure~\ref{fig:usefulness-mytool} shows the usefulness of \mytool{}'s recommendation for the \numcomment{} open issues. The results show that the issues recommended by \mytool{} are useful in helping developers to find more contexts about the bug (\nummytoolcon{}/\numcomment{}=55.56\% issues) and in repairing the bug (\nummytoolrep{}/\numcomment{}=25\% issues). These results are consistent with our findings in Section~\ref{sec:usefulsim} where similar bugs help developers in finding more context about the bug and in bug fixing.
 



\begin{framed}
\noindent\textbf{Answer to Q2:} The similar bugs recommended by \mytool{} could help developers in finding more context about the bug and fixing the bug.  
\end{framed}


\subsection{Feedback from developers}
For each of the \numcomment{} issues, we find all its relevant recommended issues (one open issues may have multiple relevant issues), and leave a comment using the format below: 
\begin{framed}
\noindent I noticed that there is a similar problem at https://xyz. Perhaps we can refer to \underline{this issue} to find more context about the bug?

\noindent Or maybe this can help us find the faulty lines and we can also refer to the \underline{fix} for the bug?''
\end{framed}

We exclude the ``Or maybe...'' part if the recommended issues do not help in debugging or fixing the bug. For each commented issue, we manually analyze the feedback. In total, we obtained \numevalreply{} replies for the commented issues with almost all positive feedback (except for one issue where the developer explained that they do not need help for resolving the issue because the pull request has already been prepared). Among the \numreplypos{} positive replies, the usefulness for \numevalconfirm{} of the recommended issues are confirmed by the developers, and another \numevalclose{} issues are closed by developers after receiving our comment. For the \numevalclose{} issues that are closed by developers after we left our comment, developers usually acknowledged our comment in the discussion. An example of such feedback is ``Closing in light of the comments from @...''.

Consider another example of positive feedback for the Notes app~\footnote{https://github.com/stefan-niedermann/nextcloud-notes/issues/847}. To fix the crash, we recommended upgrading the \texttt{com.google.android.material} dependency from 1.1.0 to 1.3.0-alpha02 by referring to the \mytool{}'s recommended issue. Although the developer acknowledged the value of our research by saying ``Nice research! Thank you!...'', he did not accept our pull request because he preferred waiting for the stable version of the \texttt{com.google.android.\\material} dependency to be released instead of upgrading it to the alpha version (1.3.0-alpha02). This example shows that \mytool{} can complement existing automated dependency update tools (e.g., dependabot) by finding the fix for a dependency-related crash before getting the official automated dependency update.



\section{Threats to validity}

\noindent\textbf{External.} Our study of similar bugs and our evaluation results may not generalize beyond the evaluated Java projects and Android apps. To mitigate this threat, we evaluate many open-source Java projects and Android apps that are popular in GitHub. Moreover, we perform our evaluation on open (unresolved) GitHub issues from real-world projects and made our data publicly available~\citep{our-web}.  


\noindent\textbf{Internal.} To reduce bias in selection of open issues, we wrote scripts to automatically crawl open GitHub issues from popular Java/Android projects. During the manual inspection and classification of the usefulness of each GitHub issue, two authors of the paper review the results independently and meet to resolve any disagreement.

\noindent\textbf{Construct.} Although many automated repair approaches are publicly available, we did not compare \mytool{} against them because to the best of our knowledge, \mytool{} is the first tool that uses the idea of \cofix{} for debugging and program repair. Moreover, we evaluate the effectiveness of \mytool{} mainly based on the relevance and the usefulness of the recommended GitHub issues but other aspects (e.g., the time taken in reading and understanding the recommended issues) could be important. We mitigate this threat by reporting all issues that we considered useful to developers, and manually analyzing developers' feedback.

\section{Related Work}

\noindent\textbf{Automated Program Repair.} There are many search-based repair techniques~\citep{rsrepair,tan2016anti,wen2018context,liu2019tbar,genprog-icse09,tan2018repairing,yuan2018arja,relifix,martinez2016astor,arcuri-CEC08,gao2019crash}. These techniques usually search for patches that are generated either based on specifically designed mutation operators or pre-defined set of repair templates. Meanwhile, there are several repair approaches that use bug reports to enhance the fault localization and bug fixing performance~\citep{r2fix,bugreportrepair}. R2Fix~\citep{r2fix} relies on several fixed bug fix patterns for fixing buffer overflows, null pointer bugs, and memory leaks. iFixR uses bug reports for IR-based fault localization, and relies on fix patterns for generating patches~\citep{bugreportrepair}. Our technique is different from existing approaches because: (1) we use bug reports and their corresponding bug fix commits from \emph{other similar bugs} instead of bug report from buggy program, (2) our approach are not limited by a fixed set of repair templates as it searches in GitHub for more than millions of patches with their corresponding issues, and (3) our goal is to recommend bug reports with bug fix commits that may contain richer information to explain the provided defect rather than generating a simple patch automatically. 

\noindent\textbf{Mining existing patches.} Similar to techniques that rely on mining or extracting recurring bug fix patterns for generating patches~\citep{prophet,le2016history,mei-et-al-ASE15,bader2019getafix,precfix,koyuncu2020fixminer,codesearch,xin2017leveraging}, \mytool{} mines patches from similar bugs. \mytool{} is also related to approaches that use the concept of similar code for finding bug fixes~\citep{jiang2018shaping,ke2015repairing,DBLP:conf/icsm/AsadGS19,DBLP:conf/icse/SahaSP19,codephage}. Among these techniques, \precfix{} that recommends patches based on mined defect-patch pairs from histories of industrial codebase~\citep{precfix}, is the most relevant to our approach. Instead of recommending patches from codebase of a single organization (Alibaba), \mytool{} recommend patches that come from up to millions of open source GitHub projects. Similar to \mytool{}, \emph{QACrashFix} generates a query with exception message to search for relevant Q\&A pages. However, \mytool{} differs from all these techniques in several aspects: (1) our study shows that similar bugs are defined by several characteristics where code similarity and similar exception are part of these characteristics; (2) to the best of our knowledge, \mytool{} is the first general approach that could repair defects that span across different types of files (build files and source files), whereas other techniques could only fix defects in specific type of files; (3) \mytool{} could help developers in tasks beyond program repair, including debugging, defect understanding, and resolving issues faster. 

\noindent\textbf{Studies on similar bugs.} Several duplicate bug reports detection approaches leverage the similarities between bug reports for recommending similar bugs \citep{sun2011towards,yang2016combining,wang2008approach,rocha2016empirical}. These techniques consider ``similar bugs'' as bugs that involve handling many common files. Similar to our approach, NextBug recommends similar bugs using the textual description of bug reports~\cite{rocha2015nextbug}. \mytool{} is different from NextBug in several key aspects: (1) NextBug is designed specifically for BugZilla where the bug component is clearly specified in a field by the person who reported the bug, whereas the information about the bug component is embedded within the text of the GitHub issue; (2) NextBug identifies similar bug reports based on bug component and textual description, whereas \mytool{} finds similar bugs by using performing several similarity analyses, including code similarity, dependency similarity, permission similarity and UI similarity; and (3) NextBug focuses on only projects within the same organization and has been evaluated only on Mozilla products, whereas \mytool{} are designed to recommend bug fixes drawn from a wide variety of open-source projects in GitHub.
Nevertheless, compared to existing duplicate bug reports detection approaches, our concept of similar bugs is more general as our definition considers several characteristics of similar bugs. Moreover, \mytool{} could identify similar bugs that occur across different projects, whereas prior definition only considers similar bugs within a particular software project.

 \noindent\textbf{Collaborative Programming.} Although \mytool{} is inspired by the concept of collaborative testing~\citep{collabbug,xie2014cooperative,long2014enabling,long2016coordinated} and pair programming~\citep{williams2000strengthening}, \mytool{} is designed with the goal of recommending GitHub issues for assisting developers in debugging and fixing rather than finding bugs. Moreover, our study of similar bugs strengthens the observation of prior research on \cofind{}. Specifically, prior study shows that similar bugs may exist across different Android apps~\citep{collabbug}, whereas our study finds that similar bugs may even exist across different Java projects. Moreover, our study shows that the concept of similar bugs could be useful for tasks beyond testing (e.g., debugging and fixing). 

\noindent\textbf{Studies on GitHub.} Prior studies on GitHub focus on the characteristics of repositories hosted on GitHub~\citep{kalliamvakou2014promises,coelho2015unveiling}, the social factors that influence the contribution in GitHub~\citep{tsay2014influence}, and its pull-based software development model~\citep{gousios2014exploratory,gousios2016work}. Different from these studies, our study focuses on the characteristics and the usefulness of similar bugs.

\noindent\textbf{Recommendation systems.} Many recommendation systems have been proposed for performing different software engineering tasks~\citep{Zhou:2012:BFM:2337223.2337226,Ye:2014:LRR:2635868.2635874,5235134,Nguyen:2011:TAN:2190078.2190181}. Some of these systems rank bug reports for fault localization~\citep{Zhou:2012:BFM:2337223.2337226,Ye:2014:LRR:2635868.2635874,Nguyen:2011:TAN:2190078.2190181}. Different from these approaches, \mytool{} recommends GitHub issues for debugging and fixing bugs for Java and Android projects.

\section{Conclusion}
We introduce \cofix{}, a concept that aims to resolve open GitHub issues by suggesting similar bugs. Our study shows that similar bugs could help developers in getting more contexts about the bug and fixing the bug. Inspired by our study, we designed \mytool{}. Given an open GitHub issue $d$, \mytool{} automatically suggests a closed GitHub issue that could be used for debugging and fixing the bug in $d$. Our evaluation on \numeval{} open GitHub issues show that \mytool{} could recommend relevant GitHub issues with similar bugs. In total, \mytool{} successfully recommended relevant issues for \numcomment{} open issues, where \numreplypos{} of these issues received positive feedback and other issues are awaiting reply from developers.

Although we currently search for open GitHub issues manually, we envision \mytool{} to be used as a bot that monitors constantly for open issues and automatically suggests issues with similar bugs. With the rising demands for software development bots~\citep{urli2018design,dependabot}, we believe that \mytool{} is a step towards this direction. In future, we would like to integrate \mytool{} with automated dependency update tools as such integration will be useful based on the developers' feedback. 





\begin{acknowledgements}
This work was supported by the National Natural Science Foundation of China (Grant No.61902170).
\end{acknowledgements}

%
%

\bibliographystyle{spbasic}      
\bibliography{bibliography}


\end{document}